# Portable Magnetic Particle Spectrometer (MPS) for Future Rapid and Wash-free Bioassays

Kai Wu[a,†,*], Vinit Kumar Chugh[a,†], Arturo di Girolamo[a], Jinming Liu[a], Renata Saha[a], Diqing Su[b], Venkatramana D. Krishna[c], Abilash Nair[a], Will Davies[d], Andrew Yongqiang Wang[e], Maxim C-J Cheeran[c,*], and Jian-Ping Wang[a,b,*]

[a]Department of Electrical and Computer Engineering, University of Minnesota, Minneapolis, MN 55455, United States

[b]Department of Chemical Engineering and Material Science, University of Minnesota, Minneapolis, MN 55455, United States

[c]Department of Veterinary Population Medicine, University of Minnesota, St. Paul, MN 55108, United States

[d]Department of Computer Science, University of Minnesota, Minneapolis, MN 55455, USA

[e]Ocean Nano Tech LLC, San Diego, CA 92126, USA

**ABSTRACT:** Nowadays, there is an increasing demand for more accessible routine diagnostics for patients with respect to high accuracy, ease of use, and low cost. However, the quantitative and high accuracy bioassays in large hospitals and laboratories usually require trained technicians and equipment that is both bulky and expensive. In addition, the multi-step bioassays and long turnaround time could severely affect the disease surveillance and control especially in pandemics such as influenza and COVID-19. In view of this, a portable, quantitative bioassay device will be valuable in regions with scarce medical resources and help relieve burden on local healthcare systems. Herein, we introduce the MagiCoil diagnostic device, an inexpensive, portable, quantitative and rapid bioassay platform based on magnetic particle spectrometer (MPS) technique. MPS detects the dynamic magnetic responses of magnetic nanoparticles (MNPs) and uses the harmonics from oscillating MNPs as metrics for sensitive and quantitative bioassays. This device does not require trained technicians to operate and employs a fully automatic, one-step, wash-free assay with user friendly smartphone interface. Using a streptavidin-biotin binding system as a model, we show that the detection limit of the current portable device for streptavidin is 64 nM (equal to 5.12 pmole). In addition, this MPS technique is very versatile and allows for the detection of different diseases just by changing the surface modifications on MNPs. It's foreseen that this kind of portable device can transform the multi-step, laboratory-based bioassays to one-step field testing in non-clinical settings such as schools, homes, offices, etc.

## 1. INTRODUCTION

The past decade has seen the continuous advancing of disease diagnostic platforms in a wide variety of research areas such as magnetic, optical, mechanical, and electrochemical sensing systems.[1–13] However, the processes of developing these platforms towards point-of-care (POC) devices for field testing are largely delayed despite their promising high sensitivity.[14–17] Most of the diagnostic platforms are complicated to use on site since they rely on expensive and/or bulky laboratory equipment as well as trained technicians to operate. Furthermore, biological samples often need to be pre-processed to remove substances such as blood cells from whole blood samples that may interfere with the fluorescence signal. These factors lead to relatively expensive diagnostics and long turnaround time. Although there are strip tests available in the market for at-home pregnancy and common diseases testing that are easy-to-use and inexpensive,[18,19] these strip tests are only limited to certain diseases and there is a big concern raised by researchers on the accuracy such as high false-positive rates.[20,21] Furthermore, the strip test results are often qualitative (or binary) which limits their capability for daily monitoring of chronic disease and in-depth disease analysis.

In recent years, the demand for high accuracy, inexpensive, and easy-to-use POC devices for routine daily diagnostics that are more accessible to patients is tremendously increasing. During the COVID-19 pandemic, the inaccessibility to portable diagnostic devices has put great pressure on the local healthcare systems especially in developing countries and rural areas.[22–25] Diagnostic platforms that combine the accessibility of strip tests and the high accuracy and quantitative nature of laboratory-based tests will greatly change current situation.

Herein, we introduce a portable, quantitative diagnostic platform based on a magnetic particle spectrometer (MPS) called MagiCoil, that can be operated by layperson in non-clinical settings such as schools, homes, and offices, etc., without much training requirements. This technique relies on detecting dynamic magnetic responses of magnetic nanoparticles (MNPs) from biological samples.[10,26–34] Since MNPs are the sole sources of magnetic signal and most biological samples are non-magnetic or paramagnetic, this MPS platform is naturally immune to the background noise from biological samples and thus, it does not require sample pre-processing and allows one-step and wash-free bioassays. Furthermore, by surface functionalizing MNPs with different ligands (e.g., carboxylic acid, amine), nucleic acids (i.e., DNA, RNA), and proteins (e.g., antibodies, streptavidin, protein A, etc.), the MNPs can be specifically modified for detecting different target analytes as well as diseases.[10,29,30,32,35,36]

## 2. EXPERIMENTAL SECTION

**2.1. MagiCoil Portable Device.** Figure 1(a) shows a photograph of the developed MagiCoil portable device along with smartphone application user interface. The overall dimensions of this device are 212 mm (L) × 84 mm (W) × 72 mm (H). It is powered by wall-plug and can communicate with smartphones (Android and IOS systems), tablets, computers through Bluetooth and USB.[37] The device shell (Figure 1(a): i) is 3D printed with polylactic acid (PLA) material. The biofluid sample holder is a flat bottom, USP type I, glass vial with dimensions of 31 mm × 5 mm and volume capacity of 0.25 mL (Figure 1(a): ii). This kind of glass vial is one-time use only and disposable and it can be seamlessly inserted into the sample loading port (Figure 1(a): iii) from the top of the MagiCoil device. The user interface (Figure 1(a): iv) gives users step-by-step instructions on carrying out the testing. Figure 1(b) shows the photograph of two circuit boards and three sets of copper coils for generating magnetic fields as well as collecting dynamic magnetic responses of MNPs. MagiCoil portable device 3D models are shown in Figure 1(c) & (d). The top circuit board (Figure 1(c): v) is the signal readout board and consists of instrumentation amplifier, bandpass filters and ADC stages. Microcontroller unit is also housed on the same board to communicate with ADC circuitry. The bottom circuit board (Figure 1(c): vi) consists of mainly the power generation ICs and coil driver circuit to generate the required magnetic fields.

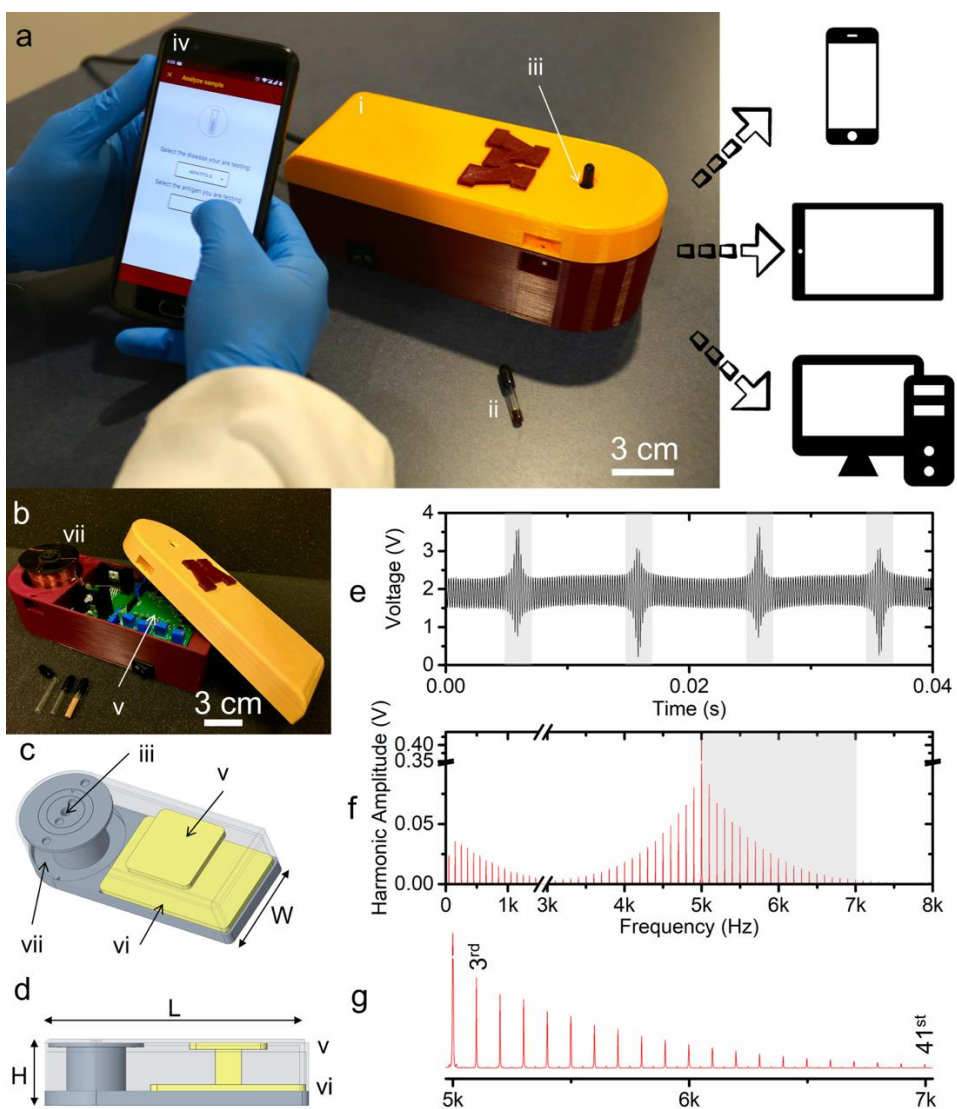

Figure 1. (a) Photograph of the MagiCoil portable device with smartphone application. The overall dimensions of device are 212 mm (L) × 84 mm (W) × 72 mm (H). (i) Device shell is 3D printed using PLA material. (ii) Disposable, USP type I glass vial containing MNP sample. (iii) Sample loading port. (iv) Smartphone application. (b) Photograph of the internal structures of MagiCoil device. (c) 3D model of MagiCoil device with (v) top and (vi) bottom circuit boards, and (vii) three sets of copper coils for generating magnetic driving fields and collecting

dynamic magnetic responses of MNPs. (d) Side view of 3D model with length (L) and height (H) labeled. (e) Discrete time voltage signal collected from pick-up coils during two periods of low frequency field. The dynamic magnetic responses of MNPs cause visible spikes as highlighted in grey regions. (f) The frequency domain MPS spectra from (e). Higher harmonics are observed. (g) is the enlarged view of higher harmonics (the 3$^{rd}$ to the 41$^{st}$ harmonic) between 5 kHz and 7 kHz. More details on the 3D models and user interfaces are given in Supporting Information S1.

**2.2. Circuit Design of MagiCoil Portable Device.** Figure 2 shows the block level breakdown of the developed MagiCoil device. One of the key requirements for MagiCoil modality for bioassay applications is the generation of phase stable magnetic fields. DDS IC from Analog Devices AD9833 is used to generate stable low frequency ($f_L$) and high frequency ($f_H$) sinusoid fields for this application. Frequencies $f_L$ and $f_H$ are kept at 50 Hz and 5 kHz, respectively. DC-shift and gain stages are implemented to obtain suitable signal amplitudes using ultra-high precision operational amplifiers OPA189 before feeding the signal to voltage source implementation using high-voltage, high-current operational amplifier from Texas Instruments (TI) OPA548. For the application presented in this work, the magnitude of low frequency field is kept at 250 Oe and that of high frequency field is kept at 25 Oe.

Differential voltage signal generated form balanced pick-up coils (Figure 3(a): iv) is amplified using instrumentation amplifier by TI INA128. Sallen-key implementation of high-pass and low-pass filters are used for signal-to-noise-ratio (SNR) improvement followed by a DC-shift stage, all implemented using the operational amplifiers OPA189. A 24-bit pseudo-differential amplifier by Linear Technology LTC2368-24 is used to sample the filtered signal. STM32F767 from STMicroelectronics is the choice of microcontroller this application enabling communication of real-time sampled data at 316 kSPS with on-board ADC.

For each vial of bioassay, the MagiCoil device records 170,000 samples which is an effective time of only 0.54 s. Real-time communication of the sampled data with laptop is handled using a custom Python script utilizing USART protocol. FTDI cable TTL-232RG is utilized to enable this communication between desktop and on-board microcontroller. The discrete time voltage signal collected from one MNP sample is shown in Figure 1(e), during two periods of low frequency field. Visible spikes due to the dynamic magnetic responses of MNPs are highlighted in grey regions. These spikes are responsible for the higher harmonics in MPS spectra as shown in Figure 1(f). Zoom in view of the 3$^{rd}$ to the 41$^{st}$ harmonics between 5 kHz and 7 kHz are also given in Figure 1(g). At current stage, the post processing of the collected discrete time voltage signal is handled by MATLAB.

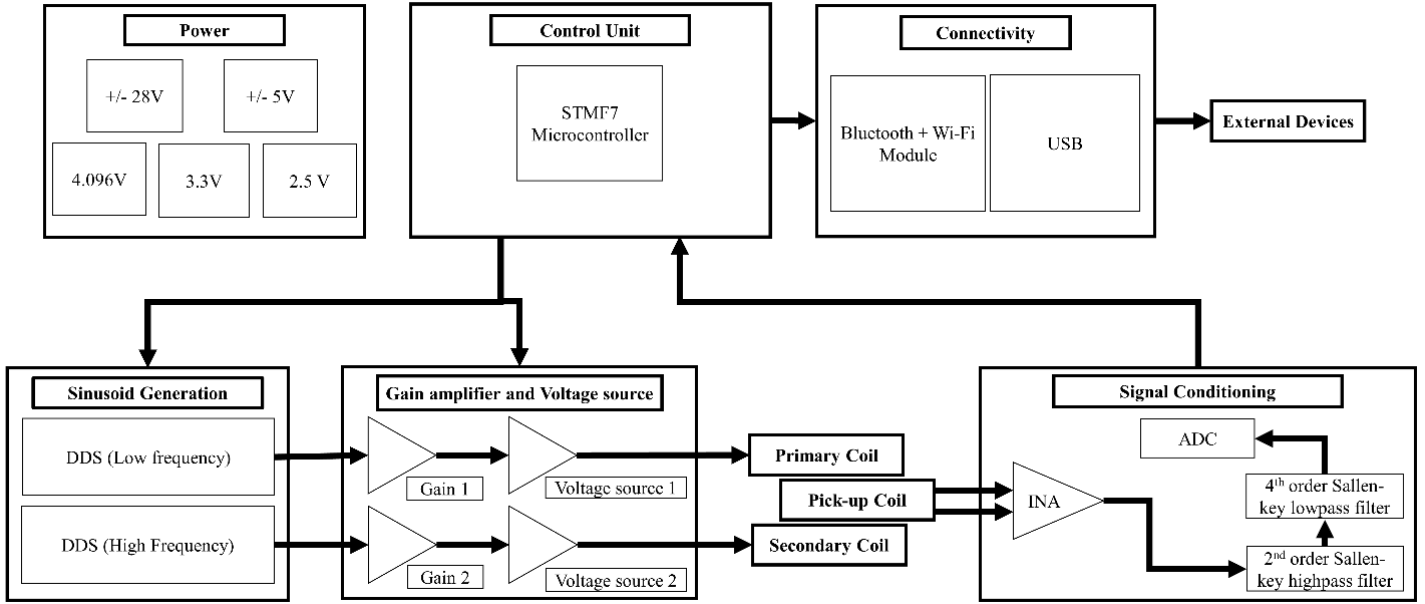

Figure 2. Block diagram of MagiCoil portable device circuit design.

**2.3. Detection Steps on MagiCoil Portable Device.** Figure 3(a) shows the general steps of carrying out one bioassay on MagiCoil portable device. Biofluid sample containing target biomolecules is dropped into a glass vial which is preloaded with a fixed amount of MNPs (Figure 3(a): i). The mixture is incubated at room temperature on plate shaker for fixed amount of time (Figure 3(a): ii) to allow the specific binding. Then the glass vial is inserted into the sample loading port for data collection (Figure 3(a): iii).

**2.4. MPS-based Bioassay Detection Mechanism.** The MagiCoil data collection part consists of three sets of copper coils: a pair of differentially wound pick-up coils (Figure 3(a): iv); one set of high frequency field $f_H$ driving coil (Figure 3(a): v); one set of low frequency field $f_L$ driving coil (Figure 3(a): vi). Figure 3(b) shows a typical MPS spectra pattern collected from samples with MNPs loaded (Figure 3(a): vii) and without MNP loaded (Figure 3(a): viii). Under the application of oscillating magnetic fields, the magnetic moments of MNPs rotate along the external magnetic field direction, this process generate dynamic magnetic responses that can be detected by the pick-up coils.[38–41] As a result, MNPs generate higher harmonics that are observed from the MPS spectra. In the results reported in this work, we only analyze the higher harmonics at $f_H+2f_L$ (the 3$^{rd}$ harmonic), $f_H+4f_L$ (the 5$^{th}$ harmonic), $f_H+6f_L$ (the 7$^{th}$ harmonic), $f_H+8f_L$ (the 9$^{th}$ harmonic), $f_H+10f_L$ (the 11$^{th}$ harmonic), $f_H+12f_L$ (the 13$^{th}$ harmonic), and $f_H+14f_L$ (the 15$^{th}$ harmonic) as highlighted in grey region from Figure 3(b). It is worth mentioning that most of the previous research work rely on the 3$^{rd}$ and the 5$^{th}$ harmonics as metrics for quantitative bioassays. Herein, we used higher order harmonics (from the 3$^{rd}$ up to the 15$^{th}$ harmonics) as reliable metrics to achieve highly sensitive and quantitative bioassay purposes.

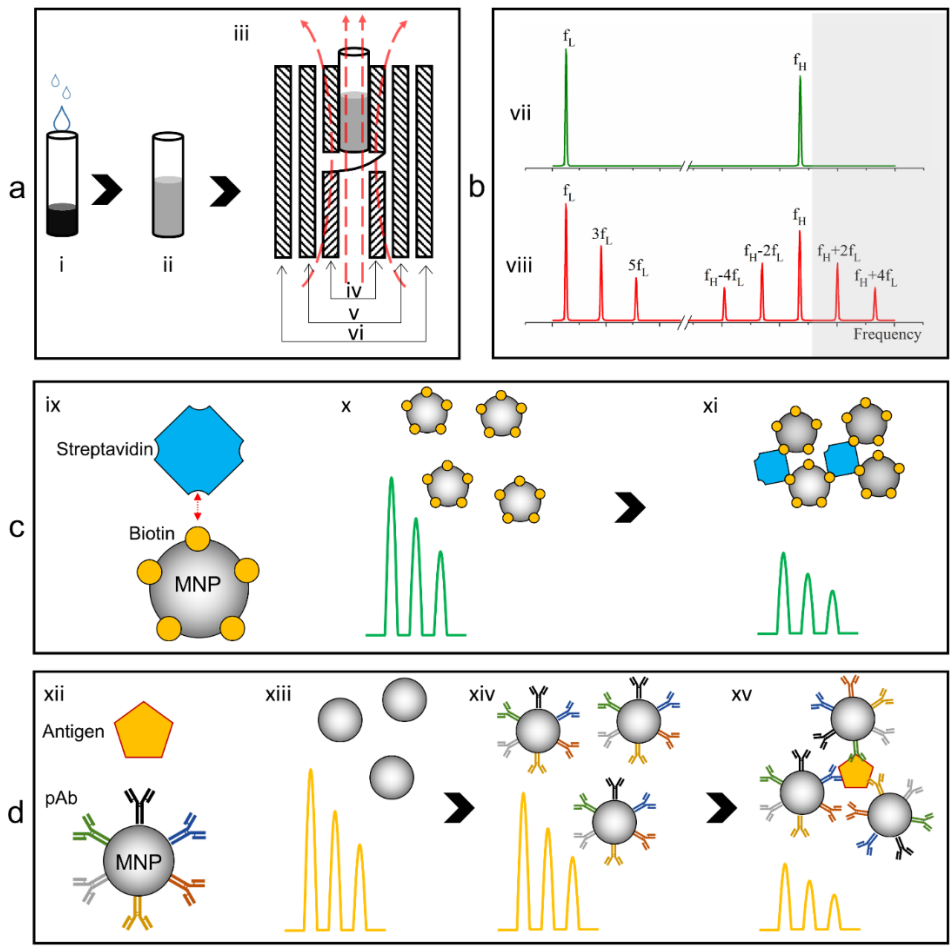

Figure 3. (a) General steps of carrying one bioassay on MagiCoil portable device. (b) MPS spectra collected from samples (vii) with MNPs loaded and (viii) without MNP loaded. (c) Schematic drawing of MPS-based biotin-streptavidin detection. (d) Schematic drawing of MPS-based antibody-antigen detection.

Taking the streptavidin detection as an example. As shown in Figure 3(c: ix), streptavidin is a homo-tetramer with an extraordinarily high affinity for biotin, one mole of streptavidin can bind with 4 moles of biotin. Well-dispersed biotinylated MNPs show high dynamic magnetic responses to external oscillating fields as well as large harmonic amplitudes (Figure 3(c): x). However, in the presence of streptavidin, biotinylated MNPs will cross-link and form clusters (Figure 3(c): xi) on streptavidin homo-tetramers. The clustering of MNPs weakens the dynamic magnetic responses and, as a result, the harmonic amplitudes drop. This difference in harmonic amplitude reduction can be used to quantitatively analyze the amount and concentration of streptavidin in the sample.

Compared to most bioassay techniques, this MPS-based bioassay does not require to remove unbound target analytes. Making it wash-free, one-step testing that is accessible by layperson in non-clinical settings.

In addition, this kind of MPS platform can be customized to detect a wide range of biomarkers as well as diseases. A more generalized detection scheme for detecting antigens using antibody-antigen interactions is given in Figure 3(d): xii). MNPs can be surface functionalized with polyclonal antibodies (pAb). In the presence of target antigens, these pAb will specifically bind to different epitopes from the antigens.[10,29] Thus, this cross-linking causes clustering of MNPs. As a result, the hydrodynamic size of MNPs gradually increases after the surface conjugation of pAb (Figure 3(d): xiii – xiv) and after the cross-linking in the presence of target antigens (Figure 3(d): xiv – xv). Similarly, this cross-linking caused MNP clustering weakens the dynamic magnetic responses and, the harmonic amplitudes drop.

## 3. MATERIALS AND METHODS

**3.1. Materials.** The iron oxide MNPs used in this work are: SHB30 and SHP30 provided by Ocean Nano Tech LLC, MP25 BN and MP25 CA purchased from Nanocs Inc. The streptavidin (product no. S4762) and phosphate-buffered saline (PBS, product no. 79378) are purchased from Sigma-Aldrich Inc. Streptavidin is a salt-free, lyophilized powder with biotin-binding capacity of 4 mol/mol (biotin), molecular weight (MW) is ~60 kDa. Sample holder is a 0.25 mL flat bottom glass vial with dimensions 31 mm × 5 mm, USP type I, manufactured by ALWSCI Technologies Co., Ltd. Round rubber end caps of 5 mm inner diameter and 15 mm height are used to seal sample holder in order to prevent liquid sample spill, manufactured by Uxcell.

**3.2. Static (dc) Hysteresis Loop Measurement.** For each MNP liquid suspension, 10 µL of sample is drawn using a pipette and dropped on a parafilm. The droplet is dried under $N_2$ at room temperature. The dc hysteresis loops of samples A – E (listed in Table 1) are measured at 300 K using a physical properties measurement system (PPMS) integrated with a vibrating sample magnetometer (Quantum Design). As plotted in Figure S2 from Supporting Information. The magnetic field is swept from -2000 Oe to +2000 Oe with a step of 2 Oe (or -300 Oe to +300 Oe with a step of 1 Oe), and the averaging time for each step is 100 ms. The dc magnetic properties of MNPs such as coercivities and saturation magnetizations are listed in Table 1.

**3.3. Particle Hydrodynamic Size Measurement.** The hydrodynamic size distribution of MNP sample is characterized using dynamic light scattering (DLS) particle tracking analyzer (Model: Microtac Nanoflex). 150 µL of samples I, III, V, VII, and X (listed in Table 2) is diluted in 1.35 mL of PBS, reaching a total sample volume of 1.5 mL mixture. This is followed by ultra-sonication for 30 minutes before the DLS characterization.

## 4. RESULTS AND DISCUSSION

**4.1. Characterization of Different Types of MNPs.** Five types of MNPs with varying sizes, concentrations, magnetic properties, and surface modifications are firstly characterized on our MagiCoil portable device. 80 µL of MNP suspension is drawn to glass vial for MPS measurements. As a comparison, the standard dc hysteresis loop measurements are carried out. From the dc hysteresis loop results in Supporting Information S2, the saturation magnetization of MNPs, $M_s$ (unit: emu/g), are ranked as: A > C > B ~ D > E. Due to the varying concentrations of MNPs in each sample, the magnetic moment per volume of suspension is calculated and listed in Table 1. Sample C with highest MNP concentration yields highest magnetic moment per microliter volume (unit: μemu/μL). The magnetic moment per volume of MNPs from highest to lowest are ranked as: C > A > B ~ D > E. All the MNP samples show negligible coercivities.

**Table 1.** Summary on Five types of iron oxide MNPs

| Sample Index | Concentration | Surface | $H_c$ (Oe)[1] | $M_s$ (2000 Oe)[1] | M (300 Oe)[1] | A3 (μV)[2] |
|---|---|---|---|---|---|---|
| **A: SHB30** | 0.8 mg/mL | Biotin | ±24.5 | 68.3 emu/g<br>54.6 μemu/μL | 59.3 emu/g<br>47.4 μemu/μL | 5067 |
| **B: SHB30** | 3 mg/mL | Biotin | ±74.1 | 33.5 emu/g<br>100.5 μemu/μL | 14.8 emu/g<br>44.4 μemu/μL | 4169 |
| **C: SHP30** | 17.5 mg/mL | Carboxylic | ±22.4 | 63.1 emu/g<br>1104.3 μemu/μL | 49.2 emu/g<br>861 μemu/μL | 81630 |
| **D: MP25 BN** | 1-2 mg/mL | Biotin | ±13.4 | 26.3-52.7 emu/g<br>26.3-105.4 μemu/μL | 14.7-29.4 emu/g<br>14.7-58.8 μemu/μL | 1924 |
| **E: MP25 CA** | 2 mg/mL | Carboxylic | ±15.9 | 17.6 emu/g<br>35.2 μemu/μL | 9.5 emu/g<br>19 μemu/μL | 1639 |

[1]VSM results are given in Supporting Information S2.

[2]MPS 3rd harmonic amplitude.

Nine independent MPS measurements are carried on each sample and the averaged harmonic amplitudes (3rd, 5th, 7th, 9th, 11th, 13th, 15th) are summarized in Figure 4(a). Harmonic amplitudes from highest to lowest are ranked as: C > A > B > D > E, which agrees with the dc hysteresis loop measurements of magnetic moment per volume of MNP suspension. Since the hydrodynamic sizes of these MNP samples are similar, the MNP amount/concentration and magnetizations directly cause the different magnetic responses and different magnitudes of higher harmonics. We listed the averaged 3rd harmonic amplitude from each sample in Table 1 for comparison. The 3rd harmonic amplitude shows similar trend with respect to the magnetic moment per volume.

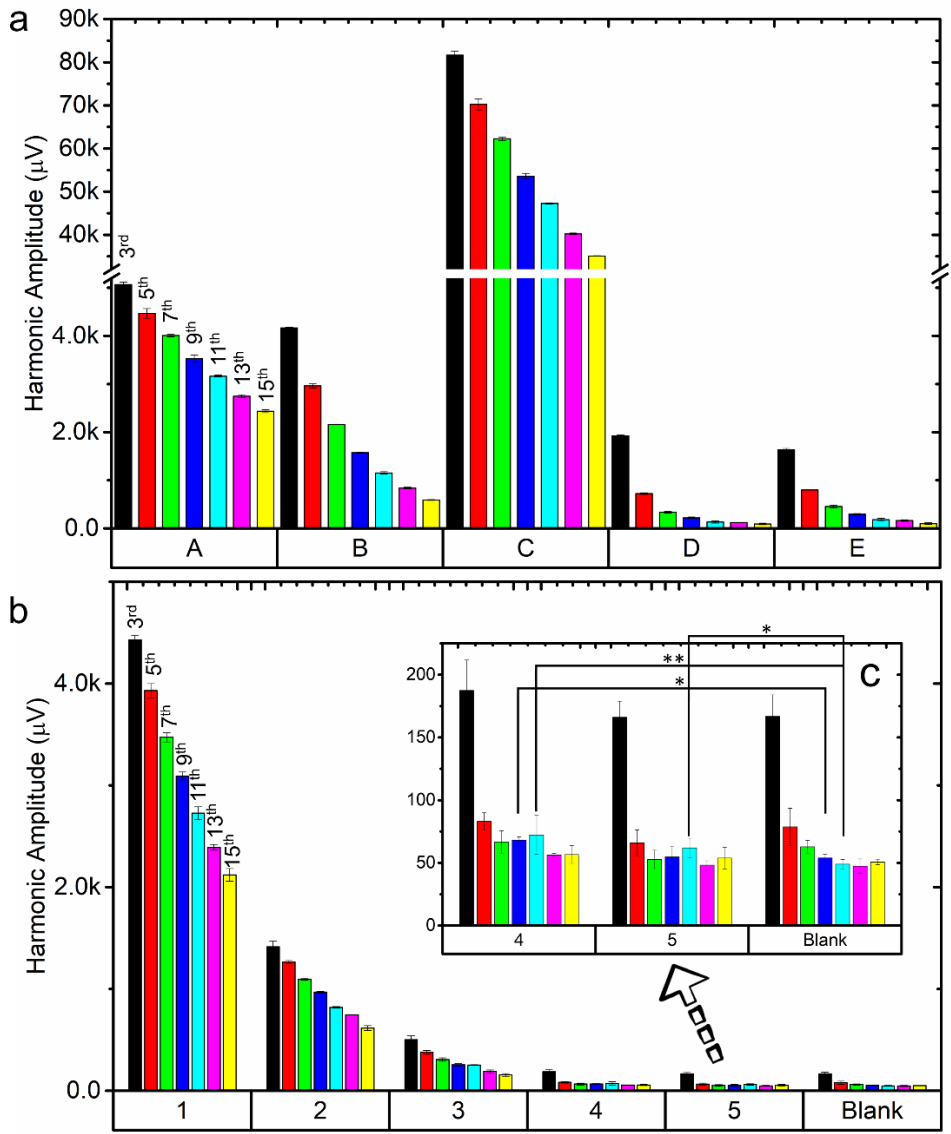

Figure 4. (a) MPS harmonics of five different types of MNPs: A - E, monitored by MagiCoil portable device. (b) MPS harmonics of varying amount of SHB30 iron oxide MNPs, monitored by MagiCoil portable device. ** $p < 0.01$; * $p < 0.05$. (c) Zoom in view of harmonics from samples 3, 4 and Blank with p-values labeled. Error bars represent standard deviations. The p-values for the $3^{rd}$ to the $15^{th}$ harmonics of samples 4 and 5 to blank sample are listed in S3 from Supporting Information.

**4.2. Minimum Amount of Iron Oxide MNPs Detectable by MagiCoil Portable Device.** To demonstrate the sensitivity of our MagiCoil portable device in detecting the lowest amount of iron oxide MNPs. 80 µL of SHB30 iron oxide MNP samples are prepared by two-fold dilutions, as listed in Table 2. From samples 1 to 5, the MNP weight per vial drops from 64 µg to 4 µg. Sample index Blank is a glass vial containing 80 µL of PBS. As shown in Figure 4(b), the higher harmonics from samples 1 – 5 are summarized and compared with the blank sample. It's clearly seen that sample 1 shows highest MPS signals, followed by samples 2 and 3. Although samples 4 and 5 show similar harmonic amplitudes compared to blank sample as shown in Figure 4(c). The two-sample t-test results show that the 9th and the 11th harmonics from sample 4 are significantly different from the blank sample, with p-values of 0.034 and 0.01, respectively. In addition, the 11th harmonic from sample 5 is significantly different from blank sample, showing a p-value of 0.04. Thus, we can conclude that our MagiCoil portable device can detect as low as 4 µg iron oxide MNPs. The calculated lowest detectable magnetic moment by this portable device is 273.2 µemu.

**Table 2.** Samples of varying amount of iron oxide MNPs.

| Sample Index | MNP Concentration (nM) | MNP Concentration (mg/mL) | MNP Weight (µg) |
|---|---|---|---|
| 1 | 27.2 | 0.8 | 64 |
| 2 | 13.6 | 0.4 | 32 |
| 3 | 6.8 | 0.2 | 16 |
| 4 | 3.4 | 0.1 | 8 |
| 5 | 1.7 | 0.05 | 4 |
| Blank | 0 | 0 | 0 |

**4.3. MagiCoil Portable Device for Streptavidin Detection.** Herein, the capability of our MagiCoil portable device for bioassay applications is demonstrated by using a streptavidin-biotin binding system. 80 µL of 1 mg/mL SHB30 iron oxide MNPs are mixed with varying concentrations/amounts of 80 µL streptavidin protein. As shown in Table 3, samples I – IX are prepared by two-fold dilution of streptavidin, with streptavidin concentrations varied from 2048 nM down to 8 nM. Sample X is prepared by mixing 80 µL of 1 mg/mL SHB30 iron oxide MNPs with 80 µL of PBS. The MNP to streptavidin ratios are also listed in Table 3. All the samples are incubated at room temperature for 30 min to allow the binding of streptavidin molecules to biotins from MNP surface.

**Table 3.** Samples of varying amount of streptavidin.

| Sample Index | Streptavidin Concentration (nM) | MNP : Streptavidin Ratio |
|---|---|---|
| I | 2048 | 1:60 |
| II | 1024 | 1:30 |
| III | 512 | 1:15 |
| IV | 256 | 1:7.5 |
| V | 128 | 1:3.75 |
| VI | 64 | 1:1.88 |
| VII | 32 | 1:0.94 |
| **VIII** | 16 | 1:0.47 |
| **IX** | 8 | 1:0.24 |
| **X** | 0 | - |

We carried out six independent MPS measurements on each sample, as shown in the scatter plots in Figure 5(a). Two-sample t-test is carried out on each sample to compare with the control group (sample index X, 0 nM). As shown in Figure 5(a), the $3^{rd}$ harmonic amplitudes from samples I - V are significantly different from the control sample X with p-values smaller than 0.05. However, the $3^{rd}$ harmonics from samples VI – IX are not significantly different from the control sample X with p-values larger than 0.05. Thus, based on the $3^{rd}$ harmonic as the sole metric, the detection limit of our MagiCoil portable device for streptavidin is 128 nM (equal to 10.24 pmole).

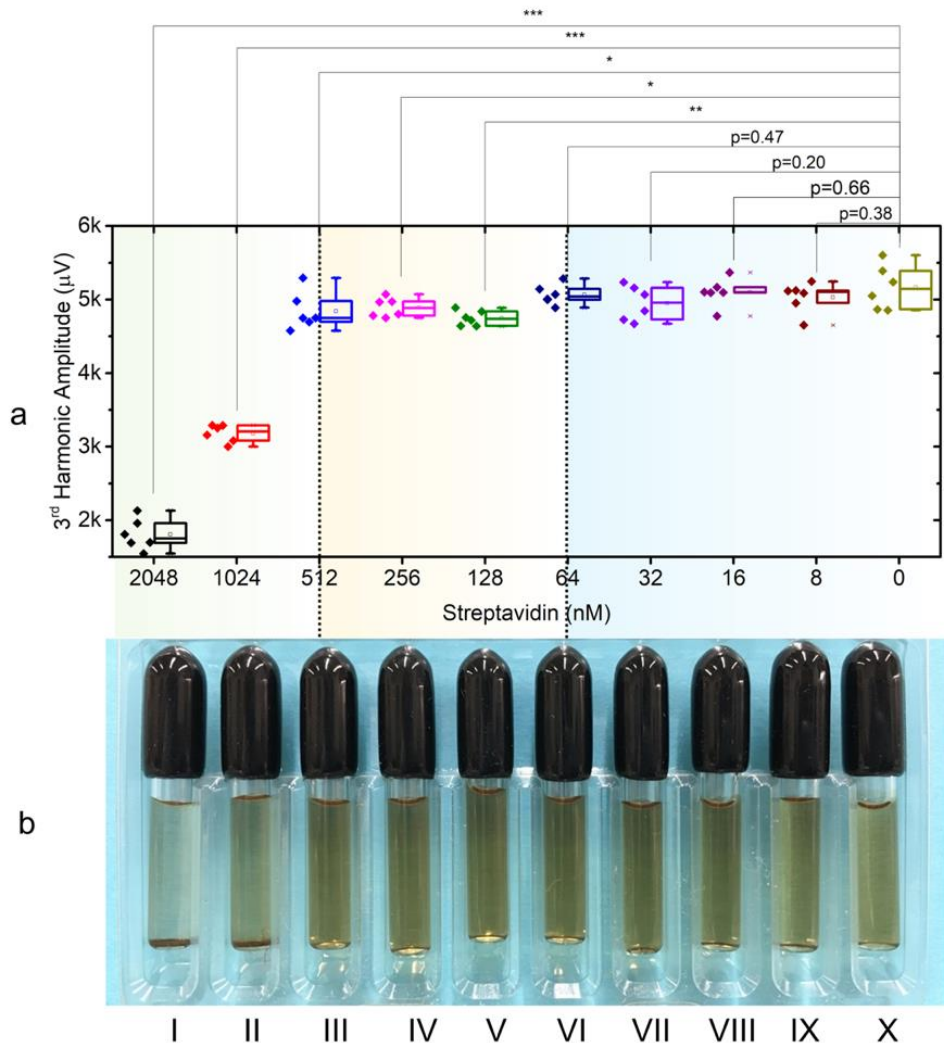

Figure 5. (a) Box plots of the $3^{rd}$ harmonics from samples I – X with varying amount of streptavidin. Two-sample t-test is carried out on each sample compared with sample X (control group). (b) Photographs of samples I – X showing different degrees of MNP clustering caused by streptavidin.

*** $p < 0.001$; ** $p < 0.01$; * $p < 0.05$. Box plots of the 5$^{th}$ and the 7$^{th}$ harmonics from samples I – X along with p-values are given in Supporting Information S4.

Figure 5(b) shows the photograph of samples I – X after the 30 min's incubation. Precipitates are observed from samples I (2048 nM) and II (1024 nM) and, the supernatants are clearer due to decreased amount of MNPs left in the solution, which indicates that larger amount of streptavidin proteins (MNP : streptavidin ratios of 1:60 and 1:30) are causing the MNPs to form clusters. Due to the cross-linking of MNPs and streptavidin, MNPs are closely compacted and their hydrodynamic sizes dramatically increase. As a result, the Brownian relaxation of bound MNPs are blocked and their dynamic magnetic responses become weaker.

The dynamic size of MNPs (and MNP clusters) from samples I, III, V, VII, and X are measured on a DLS system. As shown in Figure 6(a) & (b), samples X (0 nM, control group) and VII (32 nM) are showing very similar size distributions both with particle size peaks at ~50 nm, indicating that small amount of streptavidin in sample VII cannot effectively increase the hydrodynamic sizes of MNPs. Furthermore, the two-sample t-test on samples VII and X shows that there is no significant change of MNP hydrodynamic size (p=0.96).

As we increase the amount of streptavidin, the distribution of particle sizes becomes wider and particles with several hundreds and thousands of nanometers' sizes are observed. The two-sample t-test on samples V and X shows a p-value of 0.58 and this small difference can be explained by the peak tail in Figure 6(c) within a size range of 100 nm – 300 nm. From sample III in Figure 6(d), we observed a second particle size peak at ~175 nm. On the other hand, a second particle size peak is also observed at ~300 nm from sample I in Figure 6(e). The insets in Figure 6(d) & (e) show larger particles with several thousand nanometers size are present in samples I and III. Proving that higher concentration/amount of streptavidin is causing MNP cross-linking and clustering. As a result, the increased hydrodynamic size weakens the dynamic magnetic responses and lower MPS harmonic amplitudes are detected from MagiCoil portable device.

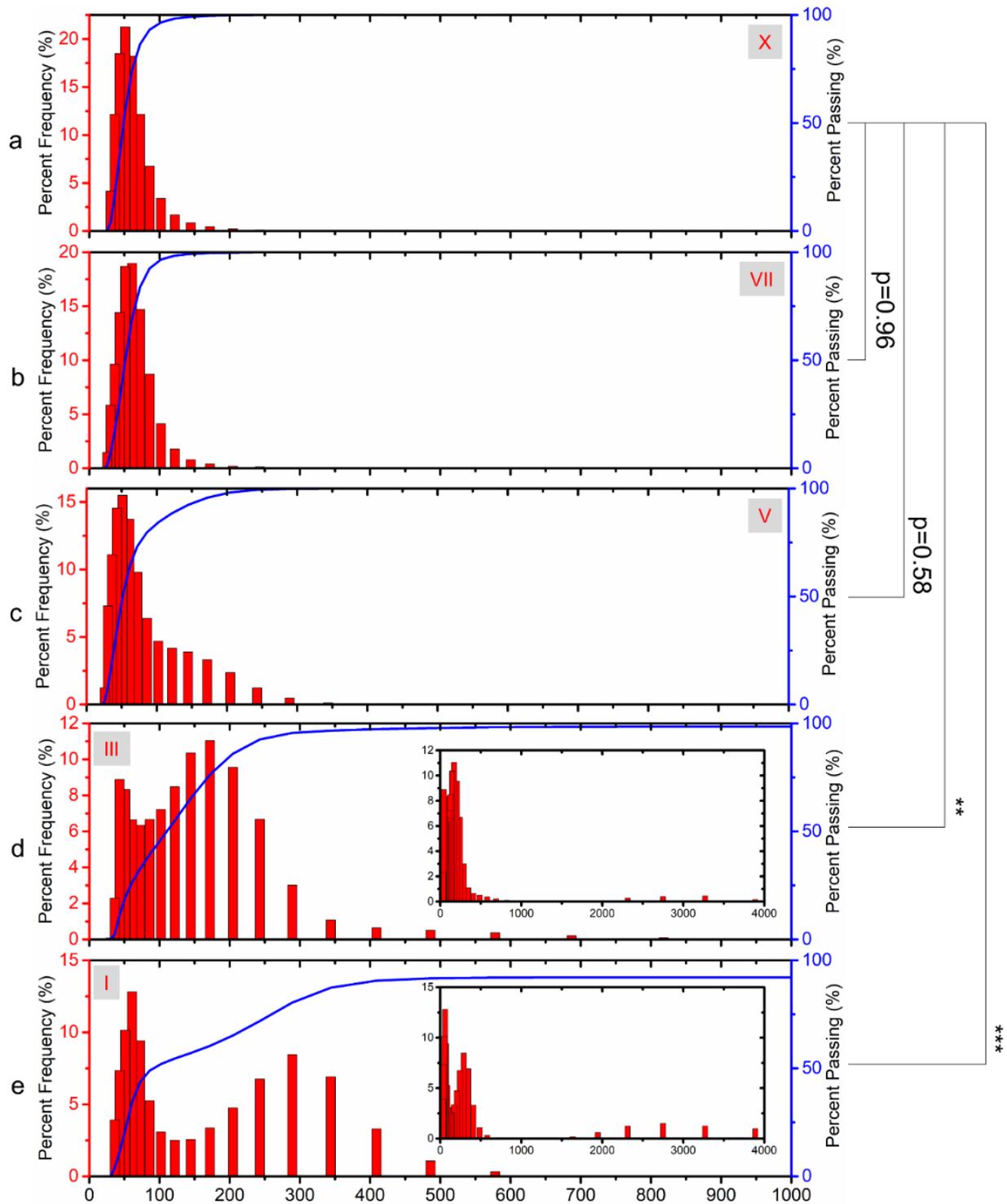

Figure 6. Hydrodynamic size distributions of MNPs from samples (a) X (control group, 0 nM), (b) VII (32 nM), (c) V (128 nM), (d) III (512 nM), and (e) I (2048 nM) measured by DLS. *** $p < 0.001$; ** $p < 0.01$; * $p < 0.05$.

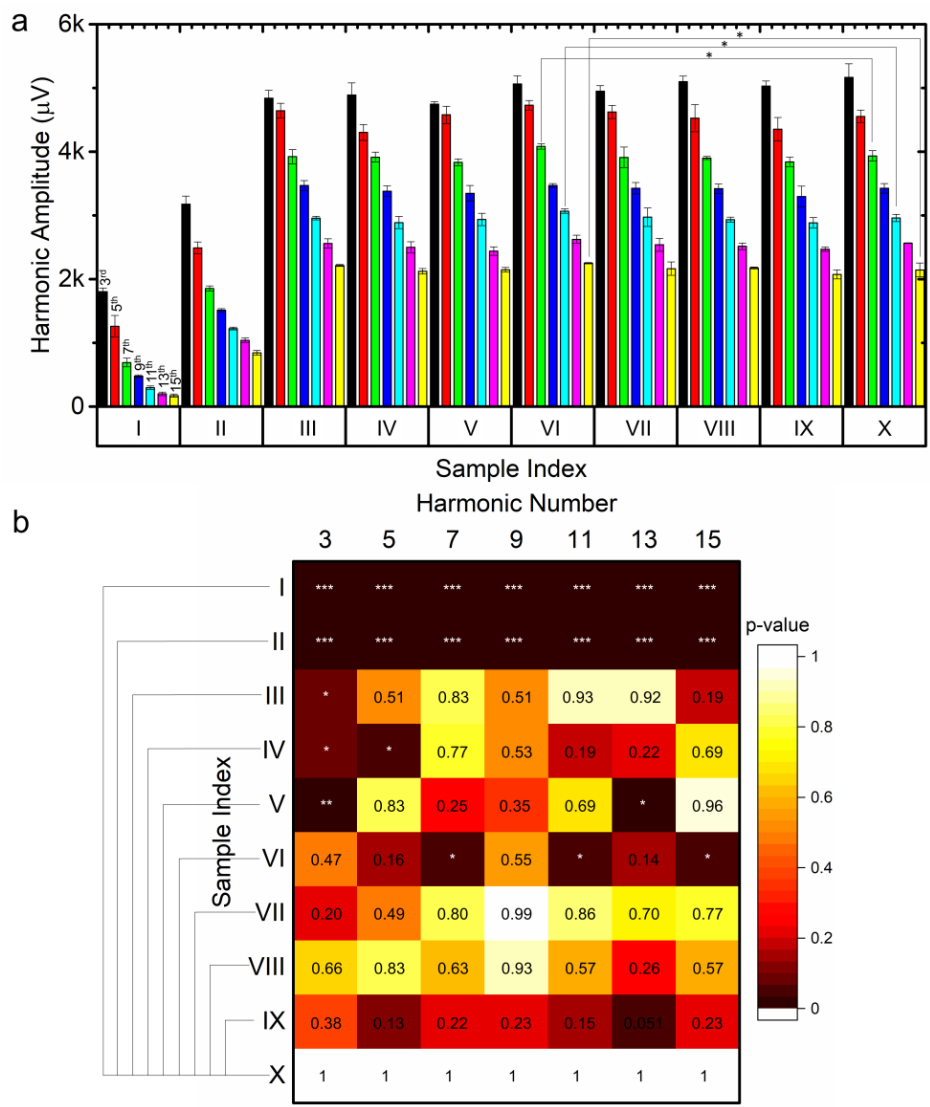

Figure 7. (a) MPS harmonics from samples I – X, monitored by MagiCoil portable device. Error bars represent standard deviations. (b) Heatmap of p-values. *** $p < 0.001$; ** $p < 0.01$; * $p < 0.05$.

Figure 7(a) summarizes the $3^{rd}$ to the $15^{th}$ harmonics from samples I to X. For each sample, harmonic amplitudes drop as harmonic index increases, but still above the noise floor. Due to the larger amount of streptavidin in the samples, samples I and II show significantly lower harmonic amplitudes than other samples. As seen from the p-value heatmap in Figure 7(b), all the higher harmonics from samples I and II are significantly different from the control sample X, with p-values smaller than 0.001. For samples III – V, the $3^{rd}$ harmonics are significantly different from control sample X. It's worth to mention that, the $7^{th}$, the $11^{th}$, and the $15^{th}$ harmonics of sample VI are significantly different from control sample X with p-values smaller than 0.05, demonstrating that by using multiple higher harmonics as metrics, our MagiCoil portable device is able to detect as low as 64 nM of streptavidin (equal to 5.12 pmole). The streptavidin concentration response curves are plotted in Supporting

Information S5 based on the 3$^{rd}$, the 5$^{th}$, and the 7$^{th}$ harmonic amplitudes from samples I to X. All the harmonic metrics show a linear dynamic range from 500 nM to 2000 nM.

## 5. CONCLUSION

Herein, we have introduced a MagiCoil portable device based on MPS technique. This portable device allows one-step, wash-free, and rapid bioassays handled by layperson in non-laboratory settings. The capability of our MagiCoil portable device is firstly demonstrated in detecting different types of MNPs. The dynamic magnetic responses of different types of MNPs are monitored by our MagiCoil device and higher harmonics are summarized. It is demonstrated that the harmonic amplitudes are directly correlated to the magnetic properties of MNPs. MNPs with higher magnetic moment per volume show stronger magnetic responses as well as higher harmonic amplitudes. Based on this, we choose high moment (high magnetization per volume) SHB30 MNPs for further studies. Then we explored the minimum amount of iron oxide MNPs that can be detected by our device. By using higher harmonics as metrics and two-sample t-test, we demonstrated that this MagiCoil portable device can detect as low as 4 µg iron oxide MNPs (equal to magnetic moment of 273.2 µemu).

We also demonstrated the feasibility of this platform for bioassay application by using the streptavidin-biotin binding system as a model. The streptavidin caused cross-linking of MNPs results in weaker dynamic magnetic responses and lower harmonic amplitudes. By analyzing the reduction of harmonic amplitudes, we can quantitatively detect the concentration/amount of target biomarkers. The streptavidin concentration response curves show a linear detection range from 500 nM to 2000 nM. Based on the 3$^{rd}$ harmonic as the sole metric, we obtained a detection limit of 128 nM (equal to 10.24 pmole). By analyzing multiple higher harmonics as metrics and using the two-sample t-test, we observed a detection limit of 64 nM for streptavidin (equal to 5.12 pmole).

To sum up, the higher harmonics from MPS spectra such as the 3$^{rd}$ to the 15$^{th}$ harmonics can be used as metrics to quantitatively characterize target biomarkers. By applying two-sample t-test on samples' higher harmonics, we can achieve better sensitivity from the MagiCoil device. In addition to the sample incubation time, the data collection step is fully automatic and raw data is transferred to laptop in 0.54 s. At current stage, the time domain discrete signal requires further processing such as discrete Fourier transform (DFT) to get frequency domain MPS spectra. In the future, this part of data processing can be combined in smartphone user applications.

Furthermore, we will extend the application of detecting different types of protein biomarkers by conjugating corresponding pAb onto MNPs. Two of the major issues leading to a sensitivity hit for current implementation of MagiCoil system are 1) Unbalanced set of coils causing for a background sinusoid which is ~2 orders higher than the response of magnetic nanoparticles; and 2) low SNR of signal at the ADC input, with present implementation we are only getting an effective number of bits (ENOB) of 12 for a 24-bit ADC which is far from ideal. The sensitivity and robustness of MagiCoil device can be further improved by designing a better set of balanced coils and implementing on-board lock-in modality for improvement of SNR. In addition, on-board microfluid channels are to be integrated on our MagiCoil device to 1) precisely control and process sub-microliter (sub-µL) biofluid samples; 2) reduce the incubation time by adding liquid mixing channels; and 3) concentrate all the MNPs in a smaller space where the external magnetic fields could be more uniform.

## ASSOCIATED CONTENT

**Supporting Information**

The supporting information is available at:

3D model of MagiCoil portable device and smartphone application user interface; Static (dc) hysteresis loops of samples A – E from Table 1; Two-sample t-test for samples 4 and 5 from Table 2 and Figure 3; Box plots of the $5^{th}$ and the $7^{th}$ harmonics from samples I – X along with p-values; Streptavidin concentration response curves.

## AUTHOR INFORMATION

**Corresponding Authors**

*E-mail: wuxx0803@umn.edu (K. W.)

*E-mail: cheeran@umn.edu (M. C-J. C)

*E-mail: jpwang@umn.edu (J.-P. W.)

**ORCID**

Kai Wu: 0000-0002-9444-6112

Vinit Kumar Chugh: 0000-0001-7818-7811

Jinming Liu: 0000-0002-4313-5816

Renata Saha: 0000-0002-0389-0083

Diqing Su: 0000-0002-5790-8744

Venkatramana D. Krishna: 0000-0002-1980-5525

Maxim C-J Cheeran: 0000-0002-5331-4746

Jian-Ping Wang: 0000-0003-2815-6624

**Author Contributions**

[†]K.W. and V.K.C. have contributed equally to this work.

**Notes**

The authors declare no conflict of interest.

## ACKNOWLEDGMENTS

This study was financially supported by the Institute of Engineering in Medicine, the Robert F. Hartmann Endowed Chair professorship, the University of Minnesota Medical School, and the University of Minnesota Physicians and Fairview Health Services through COVID-19 Rapid Response Grant. This study was also financially supported by the National Institute of Food and Agriculture (NIFA) under Award Number 2020-67021-31956. Portions of this work were conducted in the Minnesota Nano Center, which is supported by the National Science Foundation through the National Nano Coordinated Infrastructure Network (NNCI) under Award Number ECCS-1542202.

**TOC:**

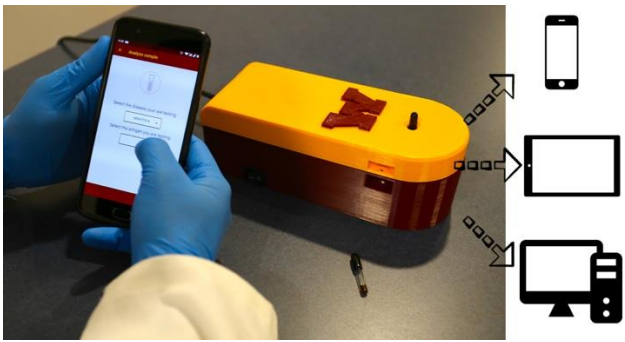

# Portable Magnetic Particle Spectrometer (MPS) for Future Rapid and Wash-free Bioassays

Kai Wu[a,†,*], Vinit Kumar Chugh[a,†], Arturo di Girolamo[a], Jinming Liu[a], Renata Saha[a], Diqing Su[b], Venkatramana D. Krishna[c], Abilash Nair[a], Will Davies[d], Andrew Yongqiang Wang[e], Maxim C-J Cheeran[c,*], and Jian-Ping Wang[a,b,*]

[a]Department of Electrical and Computer Engineering, University of Minnesota, Minneapolis, MN 55455, United States

[b]Department of Chemical Engineering and Material Science, University of Minnesota, Minneapolis, MN 55455, United States

[c]Department of Veterinary Population Medicine, University of Minnesota, St. Paul, MN 55108, United States

[d]Department of Computer Science, University of Minnesota, Minneapolis, MN 55455, USA

[e]Ocean Nano Tech LLC, San Diego, CA 92126, USA

[†]These authors have contributed equally to this work.

*E-mails: wuxx0803@umn.edu (K. W.), cheeran@umn.edu (M. C-J. C), jpwang@umn.edu (J.-P. W.)

**S1. 3D model of MagiCoil portable device and smartphone application user interface.**

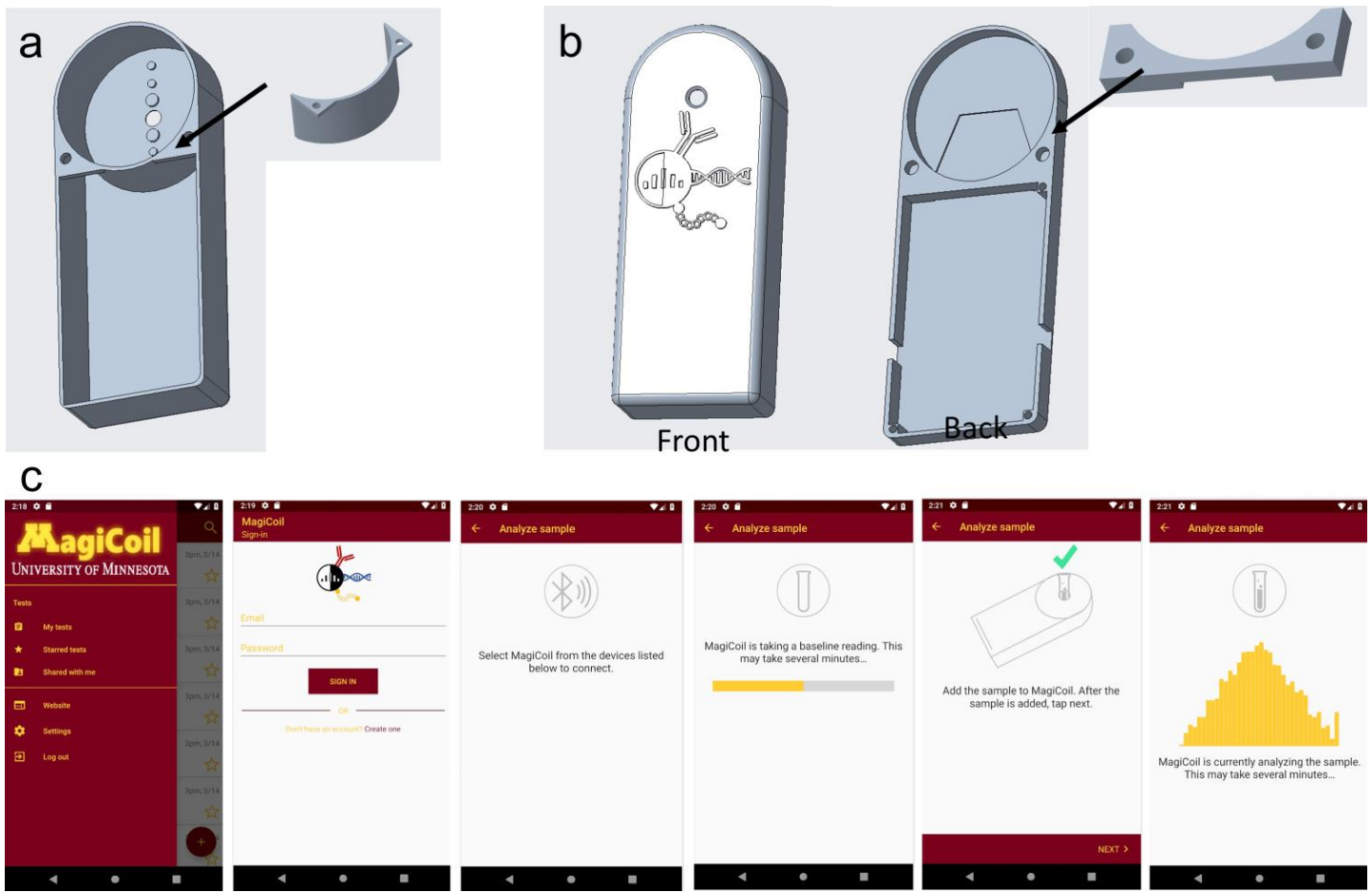

Figure S1. 3D models of the MagiCoil (a) bottom and (top) shells. (c) Screenshots of smartphone application user interfaces with instructions.

S2

## S2. Static (dc) hysteresis loops of samples A – E from Table 1.

The dc hysteresis loops of samples A – E (listed in Table 1) are measured at 300 K using a physical properties measurement system (PPMS) integrated with a vibrating sample magnetometer (Quantum Design). The magnetic field is swept from -2000 Oe to +2000 Oe with a step of 2 Oe (or -300 Oe to +300 Oe with a step of 1 Oe), and the averaging time for each step is 100 ms.

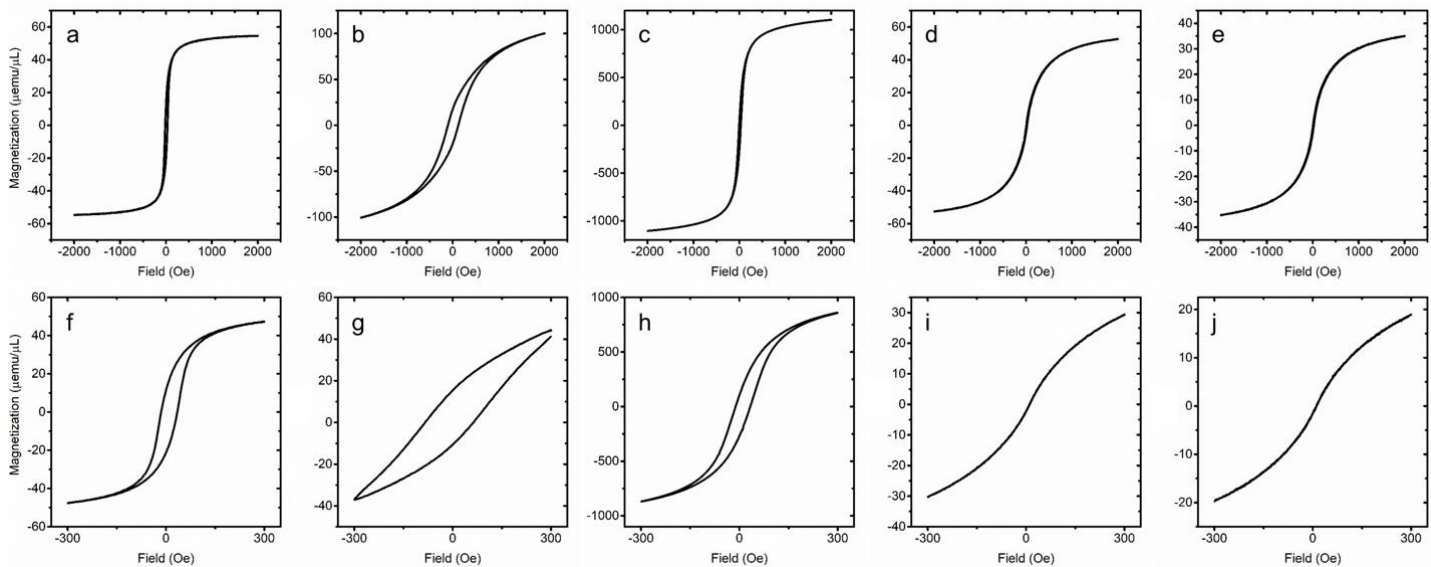

Figure S2. (a) & (f), (b) & (g), (c) & (h), (d) & (i), and (e) & (j) are the dc hysteresis loops of samples A, B, C, D, and E, respectively. In (a) – (e), the magnetic field is swept from -2000 Oe to +2000 Oe. In (f) – (j), the magnetic field is swept from -300 Oe to +300 Oe.

**S3: Two-sample t-test for samples 4 and 5 from Table 2 and Figure 3.**

Two-sample t-test is carried out to analyze the differences in harmonic amplitudes between sample 4 (sample 5) and blank sample. The p-values are listed in Table S1.

Table S1. p-values at higher harmonics.

| Harmonic # | Sample 4 with blank sample | Sample 5 with blank sample |
|---|---|---|
| 3 | 0.10 | 0.97 |
| 5 | 0.62 | 0.15 |
| 7 | 0.51 | 0.07 |
| 9 | 0.03 (*) | 0.88 |
| 11 | 0.01 (**) | 0.04 (*) |
| 13 | 0.14 | 0.91 |
| 15 | 0.25 | 0.50 |

**S4. Box plots of the 5th and the 7th harmonics from samples I – X along with p-values.**

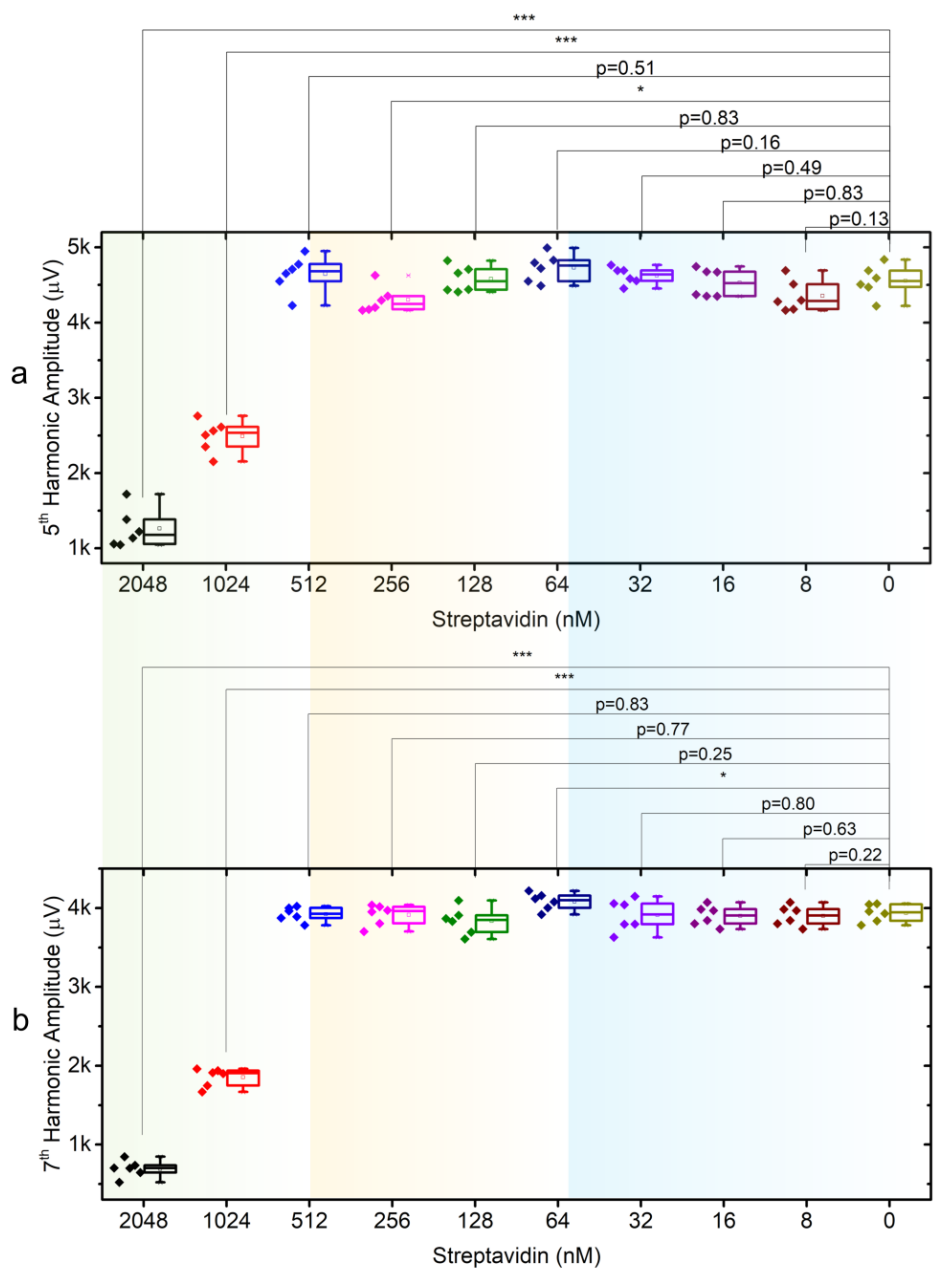

Figure S3. Box plots of the (a) 5th and the (b) 7th harmonics from samples I – X with varying amount of streptavidin. Two-sample t-test is carried out on each sample compared with the control sample X. *** $p < 0.001$; ** $p < 0.01$; * $p < 0.05$. Error bars represent standard deviations.

## S5. Streptavidin concentration response curves.

The streptavidin concentration response curves are plotted based on the 3$^{rd}$, the 5$^{th}$, and the 7$^{th}$ harmonic amplitudes from samples I to X. Detection limit is calculated by harmonic amplitudes from sample X subtracting two times of standard deviation. As shown in Figure S4, the blue dotted lines represent detection limits. Overall, all the harmonic metrics show a linear dynamic range from 500 nM to 2000 nM. The 3$^{rd}$ harmonic metric shows a detection limit of 128 nM.

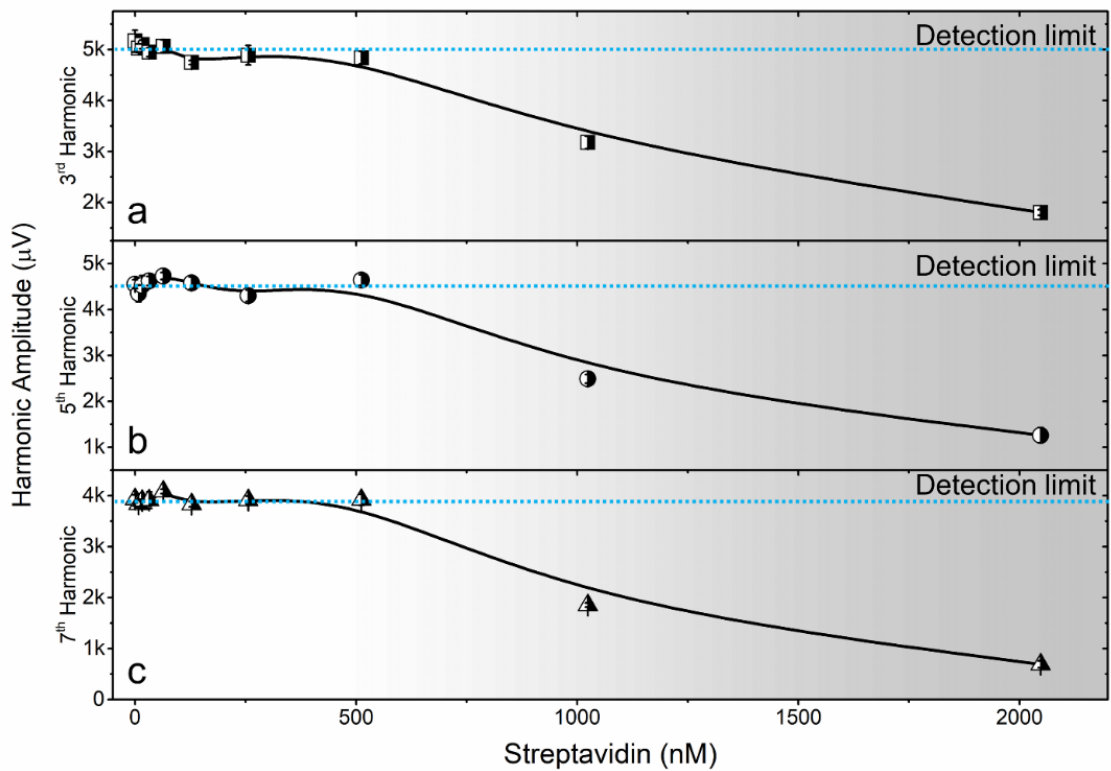

Figure S4. Streptavidin concentration response curves based on (a) the 3$^{rd}$ harmonic, (b) the 5$^{th}$ harmonic, and (c) the 7$^{th}$ harmonic. Error bars represent standard deviations.